\documentclass[a4paper,fleqn]{cas-dc}

\usepackage[authoryear]{natbib}

\usepackage{amsmath,amssymb,amsfonts}
\usepackage{algorithmic}
\usepackage{graphicx}
\usepackage{textcomp}
\usepackage{xcolor}
\usepackage{ragged2e}
\usepackage{balance}
\usepackage{tabularx}
\usepackage{colortbl}
\usepackage{multirow}
\usepackage{balance}
\usepackage{rotfloat}
\usepackage{color}

\usepackage{CJKutf8}

\usepackage[colorlinks]{hyperref}

\def\tsc#1{\csdef{#1}{\textsc{\lowercase{#1}}\xspace}}
\tsc{WGM}
\tsc{QE}
\tsc{EP}
\tsc{PMS}
\tsc{BEC}
\tsc{DE}

\begin{document}
\begin{sloppypar}

\let\WriteBookmarks\relax
\def\floatpagepagefraction{1}
\def\textpagefraction{.001}

\shorttitle{Technical Debt Management in OSS Projects}
\shortauthors{Z Li et al.}
\title [mode = title]{Technical Debt Management in OSS Projects: An Empirical Study on GitHub}

\author[1]{Zengyang Li}
\ead{zengyangli@ccnu.edu.cn}

\credit{Conceptualization of this study, Methodology, Investigation, Data curation, Writing - Original draft preparation}

\address[1]{School of Computer Science \& Hubei Provincial Key Laboratory of Artificial Intelligence and Smart Learning, \\Central China Normal University, Wuhan, China \\}

\author[1]{Yilin Peng}
\ead{pyl666@mails.ccnu.edu.cn}
\credit{Methodology, Investigation, Data curation, Software, Writing - Original draft preparation}

\author[2]{Peng Liang}
\cormark[1]
\ead{liangp@whu.edu.cn}
\credit{Conceptualization of this study, Methodology, Writing - Original draft preparation}
\address[2]{School of Computer Science, Wuhan University, Wuhan, China}

\author[3]{Apostolos Ampatzoglou}
\ead{a.ampatzoglou@uom.edu.gr}
\credit{Conceptualization of this study, Methodology}
\address[3]{Department of Applied Informatics, University of Macedonia, Thessaloniki, Greece}

\author[1]{Ran Mo}
\ead{moran@ccnu.edu.cn}
\credit{Methodology, Writing}

\author[4]{Hui Liu}
\ead{hliu@hust.edu.cn}
\credit{Conceptualization of this study, Methodology}
\address[4]{School of Artificial Intelligence and Automation, Huazhong University of Science and Technology, Wuhan, China}

\author[1]{Xiaoxiao Qi}
\ead{qixiaoxiao@mails.ccnu.edu.cn}
\credit{Investigation, Data curation}

\cortext[cor1]{Corresponding author.}


\begin{abstract}
Technical debt (TD) refers to delayed tasks and immature artifacts that may bring short-term benefits but incur extra costs of change during maintenance and evolution in the long term. TD has been extensively studied in the past decade, and numerous open source software (OSS) projects were used to explore specific aspects of TD and validate various approaches for TD management (TDM). However, there still lacks a comprehensive understanding on the practice of TDM in OSS development, which penetrates the OSS community's perception of the TD concept and how TD is managed in OSS development. To this end, we conducted an empirical study on the whole GitHub to explore the adoption and execution of TDM based on issues in OSS projects. We collected 35,278 issues labeled as TD (TD issues) distributed over 3,598 repositories in total from the issue tracking system of GitHub between 2009 and 2020. The findings are that: (1) the OSS community is embracing the TD concept; (2) the analysis of TD instances shows that TD may affect both internal and external quality of software systems; (3) only one TD issue was identified in 31.1\% of the repositories and all TD issues were identified by only one developer in 69.0\% of the repositories; 
(4) TDM was ignored in 27.3\% of the repositories after TD issues were identified; and (5) among the repositories with TD labels, 32.9\% have abandoned TDM while only 8.2\% adopt TDM as a consistent practice.
These findings provide valuable insights for practitioners in TDM and promising research directions for further investigation.\\
\end{abstract}


\maketitle

\section{Introduction}\label{chap:intro}
Technical debt (TD) refers to delayed tasks and immature artifacts that constitute a “debt” since they may bring short-term benefits but incur extra costs of change during maintenance and evolution in the long term \citep{Wa1992,AvKrOzSe2016}. TD can be classified into several types from the perspective of different stages of the software development lifecycle, such as architecture TD and design TD \citep{TOM2013, LiAvLi2015}. Due to its significant impact on software quality, TD has been a hot topic attracting great attention from both academia and industry in the last decade \citep{LiAvLi2015, Alves2016, TuPaBaOlPeDePo2017, Kruchten2019, CiLeMa2021, ToMaSj2021, Ernst2021, ViCoZaFa2022} 
, and many open source software (OSS) projects were used to explore various aspects of TD \citep{TsKeSiCh2020, LiYuLiMoYa2020, Xiao2021, TaFeAvLu2021, YuFaTUMe2022}. 

However, there still lacks a comprehensive understanding on TD management (TDM) in OSS development, which penetrates the OSS community's perception of the TD concept and how TD is managed in OSS development.
On GitHub, a proportion of issues that are explicitly tagged as TD, which are called as TD issues in this paper. 
Every TD issue is intentionally tagged with a TD label by a practitioner, which means that the practitioner actively performs TD identification, a core activity of TDM \citep{LiAvLi2015}. Since TD issues are identified by the practitioners themselves and no biases of researchers are introduced, TD issues can directly reflect the practitioner's perception and management strategy of TD. TD issues are essentially a type of self-admitted technical debt (SATD) \citep{DaKr2017,XaFeBrVa2020}. Therefore,
TD issues can be used as an important data source to comprehensively investigate TDM in OSS development.


The \textbf{goal of this work} is to explore TDM in OSS development based on TD issues. Specifically, we aim to investigate the trend of the adoption of TDM, the practitioners' perception on the TD concept, and how practitioners manage TD in OSS development. To this end, we conducted a large-scale empirical study on the TD issues of all OSS projects hosted on GitHub.

The \textbf{main contributions} of this work are summarized as follows:
  (1) To our knowledge, this work is the first study that took all repositories hosted on GitHub as the search space to understand the state of TDM in practice in a large-scale OSS ecosystem.
  (2) We performed a large-scale empirical study on TDM based on 35,278 issues with TD labels (i.e., TD issues) distributed over 3,598 repositories.
  (3) We explored TD issues from the perspectives of the involved people and management process.
  (4) We found that the practitioners' awareness of TDM is increasing over the past decade given the relatively high growth rates of new TD issues and new repositories adopting TDM.
  (5) We found that only 298 (8.2\%) repositories adopt TDM as a consistent practice, while 1,187 (32.9\%) repositories have abandoned TDM. 
  (6) We found that the power-law characteristic exists in the distributions of the proportion of repositories with TD labels over the number of TD issues and of the proportion of resolved TD issues over their open time.

The remaining of this paper is organized as follows. Section \ref{chap:relat} discusses the related work; Section \ref{chap:case} describes the study design; Section \ref{chap:results} presents the study results; Section \ref{chap:discussion} interprets the study results with the implications; Section \ref{chap:threats} highlights the threats to the validity of the results; and Section \ref{chap:conclusions} concludes this work with future research directions.

\section{Related Work}\label{chap:relat}

\subsection{Technical Debt in OSS}\label{RelatedWork_A}
Many studies investigated TD in OSS from diverse perspectives. 
Digkas \emph{et al}. looked into the evolution of TD in 66 Java projects from the Apache ecosystem \citep{DiLu2017}. They studied three aspects, \textit{i.e.}, the evolution of normalized TD, the most common types of TD, and the most costly TD. Tan \textit{et al}. conducted a case study also on the Apache ecosystem to investigate the evolution of TD remediation in Python \citep{TaFeAvLu2021}. 
Li \emph{et al}. performed an empirical study on 59 Apache OSS projects to investigate the characteristics of the interest of defect TD \citep{LiYuLiMoYa2020}. Tsoukalas \emph{et al}. carried out an empirical study on more than 100 open-source repositories from multiple OSS platforms for TD forecasting \citep{TsKeSiCh2020}.
Lenarduzzi \emph{et al}. performed a case study of 33 Apache projects to explore the spread of TD and how quickly it can be fixed \citep{LeSaTa2019}. 
Digkas \emph{et al}. conducted a case study of 27 Apache projects to investigate the stability of the introduction of TD, and the correlation between the introduction of TD and the workload of the development team \citep{GeApAlPa2020}.

The aforementioned studies took specific OSS projects as research objects to explore certain phenomena with respect to TD. In contrast, our work takes all repositories hosted on GitHub as our search space, and much more repositories with TD issues were included in our dataset.

\subsection{Technical Debt Management}\label{RelatedWork_B}

\textbf{Literature reviews on technical debt management.} A number of literature reviews have been performed to study TDM as a whole or some specific aspects of TDM.
Li \emph{et al}. conducted a systematic mapping study to examine the current state of TDM, including categories of TD, activities of TDM, approaches and tools for TDM, and challenges to TDM \citep{LiAvLi2015}. Becker \textit{et al}. performed a systematic literature review on trade-off decisions across time in TDM \citep{BeChBeMc2018}.
Lenarduzzi \emph{et al}. published a systematic literature review to understand the priority of refactoring TD over the lifespan of software, compared with developing features and fixing bugs \citep{LeBeTaMaFo2021}.
In the works mentioned above, the state of TDM was studied based on literature, while in our work we investigated the state of TDM based on practices in OSS repositories.


\textbf{Management of self-admitted technical debt.} In recent years, intensive efforts have been invested to the research of a special kind of TD -- self-admitted technical debt (SATD). 
Potdar and Shihab coined the concept of SATD, which is sub-optimal solutions intentionally introduced (\textit{e.g.}, temporary fixes) and explicitly documented using code comments \citep{AnEm2014,GiEmYa2019}. 
Most studies on SATD investigated TD admitted in the comments of source code. 
Huang \emph{et al}. proposed a model to predict whether a comment in software contains TD \citep{HuEmXiDaLi2018}. Ren \emph{et al}. presented a CNN-based approach to detect SATD in code comments \citep{ReXiXiDaWaJo2019}. Yu \emph{et al}. proposed Jitterbug to automatically identify SATD \citep{YuFaTUMe2022}.
Bavota \emph{et al}. analyzed the diffusion, evolution, participants, and impacted quality of SATD based on commits and comments of projects \citep{BaRu2016}, and their study facilitates the evaluation of the impact of TD in TDM. 
Codabux \emph{et al}. explored the types and characteristics of TD in software systems written in R language by manually analyzing the comments in the peer-review documentation of R-language OSS packages \citep{CoViFa2021}.
Zampetti \textit{et al}. surveyed industrial developers and OSS developers to compare the practices on SATD in industry and OSS development \citep{FiGiSeDi2021}. They found that both industrial and OSS developers share similar perceptions and actions regarding TD \citep{FiGiSeDi2021}. Besides, industry developers were more reluctant to acknowledge TD than OSS developers. Unlike the aforementioned works that investigate SATD and its management based on code comments, our work studies TD and TDM based on issues.

\textbf{Technical debt management based on issues.}
There are also a few studies that used issues in issue tracking systems as a data source to investigate TD and TDM.
Bellomo \emph{et al}. manually examined 1,264 issues from four software projects and identified 109 issues as TD items, which were used as the dataset to explore characteristics of TD \citep{BeNoOzPo2016}. 
Dai and Kruchten manually checked all issues of a commercial software project and identified 331 issues containing TD, and then used natural language processing and machine learning to automatically detect TD issues of the same project \citep{DaKr2017}.
Li \textit{et al}. regarded manually-identified issues containing TD in issue tracking systems as SATD, and further investigated the identification and remediation of such TD issues \citep{LiSoAv2020}. In contrast, Xavier \emph{et al}. considered issues with TD labels (\textit{i.e.}, TD issues) as a kind of SATD, and looked into why TD issues are introduced and paid \citep{XaFeBrVa2020}. Similar to Xavier \emph{et al}. \citep{XaFeBrVa2020}, in this work we only investigated issues that were tagged with TD labels (TD issues) in issue tracking systems. 

\section{Study Design}\label{chap:case}


\subsection{Objective and Research Questions}\label{DesignRQ}
The objective of this study, described using the Goal-Question-Metric (GQM) approach \citep{Ba1992}, is: to \emph{analyze} software issues tagged as TD \emph{for the purpose of} exploration \emph{with respect to} the current state of TDM in practice \emph{from the point of view of} software practitioners \emph{in the context of} OSS development. Based on the aforementioned goal, we formulated five research questions (RQs), which are described as follows: 

\begin{itemize}
\setlength{\itemsep}{0pt}
\setlength{\parsep}{0pt}
\item[\textbf{RQ1:}] \textbf{How popular do software repositories adopt TDM?} \\
\textbf{Rationale}: With this RQ, we investigated the popularity of the adoption of TDM in OSS development, which can reflect the degree of acceptance of the TD concept and practitioners' awareness of TDM in the context OSS development. 
\item[\textbf{RQ2:}] \textbf{What is the understanding of practitioners on the TD concept?}\\
\textbf{Rationale:} Practitioners understand the TD concept in different ways, which may influence the strategies and activities of TDM. The answer to this RQ is helpful to understand and improve TDM strategies.
\item[\textbf{RQ3:}] \textbf{How are TD issues identified?}\\
\textbf{Rationale:} TD identification is the first step of TDM. With this RQ, we investigated when and by whom TD issues are identified. The answer to this RQ enables us to evaluate practitioners' awareness of TDM and existing strategies of TDM.
\item[\textbf{RQ4:}] \textbf{How are TD issues resolved?} \\
\textbf{Rationale:} With this RQ, we investigated how TD issues are handled. In particular, the open time, characteristics, and proportion of reopens of resolved TD issues can provide us with multiple perspectives to understand the resolution process of TD issues. 
\item[\textbf{RQ5:}] \textbf{What is the continuity of adopting TDM for the repositories with TD labels?}\\
\textbf{Rationale:} This RQ investigates whether TDM has been continuously employed for the repositories with TD labels, and to further explore the different characteristics between repositories with different levels of continuity of adopting TDM.
\end{itemize}

\subsection{Data Collection}\label{DataCollection}



\subsubsection{Data items to be collected}
\begin{figure*}[htpb]
    \centering
    \includegraphics[width=15.0cm]{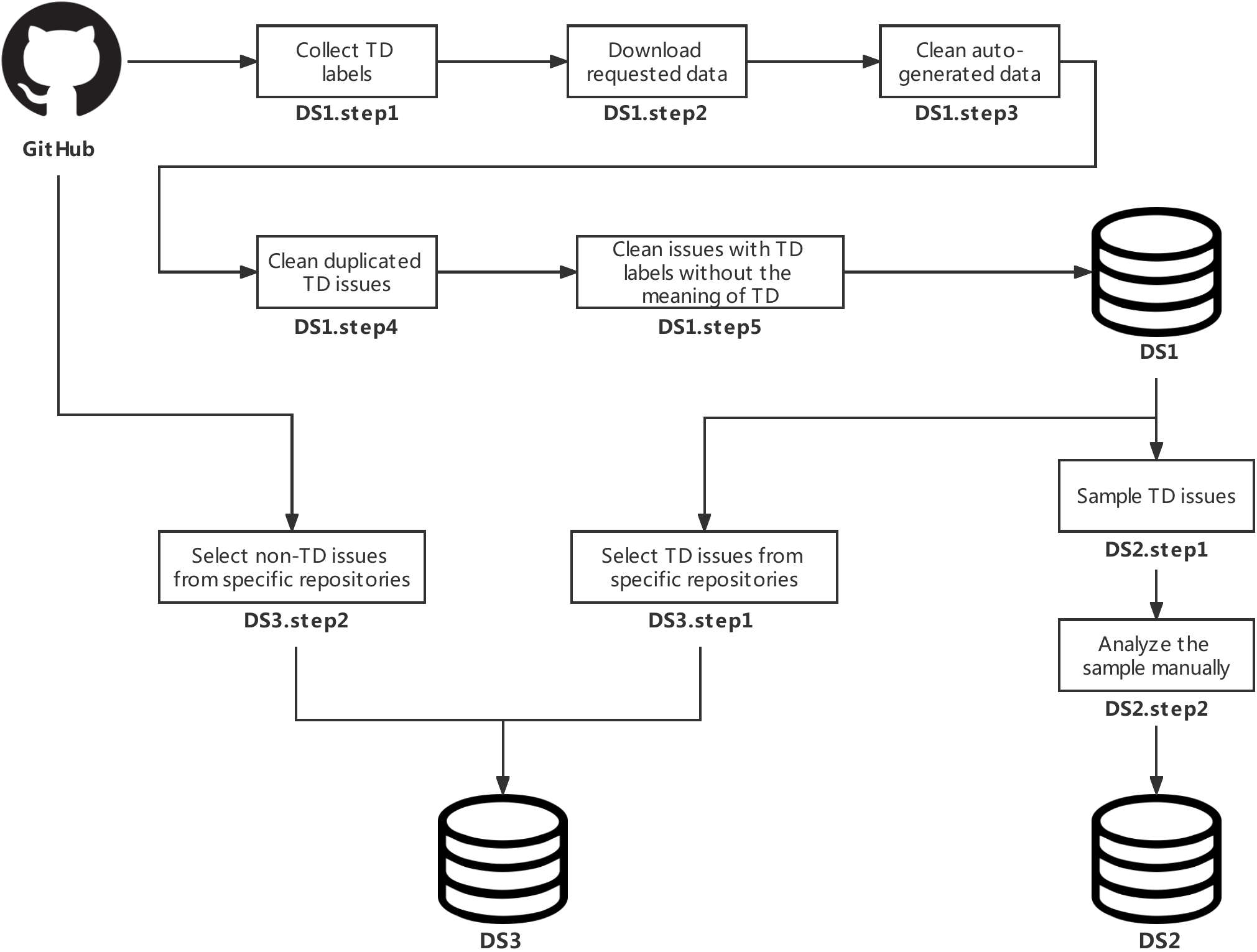}
    \caption{Procedure of data collection.}
    \label{fig:datacollectprocess}
\end{figure*}

First, to answer the five RQs formulated in Section \ref{DesignRQ}, we needed to collect all TD issues on the whole GitHub. Such TD issues are included in dataset DS1, which is the main dataset of this work. Second, to answer RQ2, besides DS1, we also needed to collect additional data items regarding the quality attributes (QAs) affected by and specific TD types of TD issues. Since manual analysis of the affected QAs and TD types of TD issues requires considerable effort, we analyzed a representative subset of DS1, which is denoted as DS2. Besides, in order to get a deeper understanding of the features of TD issues, we needed to collect data on TD issues and non-TD issues of repositories that each includes a specific amount of TD issues. Such data constitute another dataset, which is denoted as DS3. The data items to be collected are listed in Table \ref{table:dataitem}, which also provides the information on the dataset(s) containing each data item. Table \ref{table:rq&ds} shows the dataset(s) and data items used by each RQ. In this study, we collected TD issues that were reported between Jan. 1st, 2009 and Dec. 31st, 2020.
Extra data items on the whole GitHub were also collected (Section \ref{chart:extra data}). The data collection processes for the three datasets are shown in Figure \ref{fig:datacollectprocess}, in which the data collection process for the extra data is not included.

\begin{table*}[]
\caption{Data items to be collected for each issue.}
\centering
\scalebox{1.00}{
\begin{tabular}{|p{0.05\columnwidth}|p{0.23\columnwidth}|p{1.30\columnwidth}|p{0.25\columnwidth}|}
\hline
\textbf{\#} & \textbf{Name} & \textbf{Description}                                                     & \textbf{Dataset}\\ \hline
D1      & IssueID           & The global identity number of the issue on GitHub.                       & DS1, DS2, DS3 \\ \hline
D2      & Reporter          & The GitHub user identifier of the person who reported the issue.         & DS1, DS3 \\ \hline
D3      & RepTime           & The time when the issue was reported.                                    & DS1, DS3 \\ \hline
D4      & ResTime           & The time when the issue was resolved.                                    & DS1, DS3 \\ \hline
D5      & Labels            & All labels assigned to the issue.                                        & DS1, DS3 \\ \hline
D6      & IsTD              & Whether the issue was tagged with a TD label.                            & DS3\\ \hline
D7      & LabeledEvent      & The labeled event of a given issue. Each labeled event contains information such as the participant who tagged this issue, the string of label, the time of creating labeled event.                                                             & DS1\\ \hline
D8      & Repo              & The repository to which the issue belongs.                               & DS1, DS3\\ \hline
D9      & QA                & The quality attribute affected by the issue.                             & DS2\\ \hline
D10     & TDType            & The TD type of a TD issue.                                               & DS2\\ \hline
D11     & IsResolved        & Whether the issue is resolved.                                           & DS1, DS3\\ \hline
D12     & CmntEvent         & The comment event of a given issue. Each comment event contains information such as the participant who added this comment, the comment content, the time of creating this event.                                                              & DS3 \\ \hline
D13     & RepoRelation      & The relationships between issue reporter and repository.                 & DS1 \\ \hline
D14     & IssueTitle        & The text of the issue title.                                             & DS1, DS3 \\ \hline
D15     & IssueDesc         & The text of the issue description.                                       & DS1, DS3 \\ \hline
D16     & ReopenedEvent     & The reopened event of a given issue. Each reopen event contains information such as the participant who reopened issue to be resolved, the time of creating reopened event.                                                                   & DS3 \\ \hline
\end{tabular}}
\label{table:dataitem}
\end{table*}

\begin{table}[]
\centering
\caption{The dataset(s) and data items used by each RQ.}
\scalebox{1.00}{
\begin{tabular}{p{0.05\columnwidth}p{0.80\columnwidth}}
\hline
\textbf{RQ}         &\textbf{Dataset[items]} \\ \hline
RQ1                 & DS1[D1, D3, D8]\\ 
RQ2                 & DS1[D5]; DS2[D1, D9, D10]\\ 
RQ3                 & DS1[D1, D2, D3, D6, D7, D8, D14, D13, D15]; DS3[D6, D14, D15]\\ 
RQ4                 & DS1[D1, D3, D4, D8, D11]; DS3[D1, D3, D4, D6, D12, D14, D15, D16] \\
RQ5                 & DS1[D1, D2, D3, D4, D6, D8, D11, D12]\\ \hline
\end{tabular}}
\label{table:rq&ds}
\end{table}

\subsubsection{Data collection for DS1}
The data collection procedure for dataset DS1 is composed of the following 5 steps:

\textbf{Step1: Collect TD labels.}
In GitHub, developers can use a TD label (\textit{e.g.}, \textit{tech-debt}) to explicitly tag an issue as TD. To collect a complete list of candidate TD labels, we first came up with a set of equivalents (\textit{e.g.}, case conversion and word abbreviation) to "technical debt". Then, we used "debt" as a keyword to search on the issue tracking system of GitHub, and manually checked the returned issues to collect more candidate TD labels. We could not find more new TD labels after we read around the top 300 returned issues and stopped the reading process after we read the top 500 returned issues.
Finally, we added labels of TD types according to the classification of TD in \citep{LiAvLi2015}. The final set of candidate TD labels to search TD issues are provided in Table \ref{table:labelsetting}, 
where the case of the TD labels is ignored.

\textbf{Step 2: Download requested data.}
In this step, we downloaded TD issues and related data (\textit{e.g.}, labels) according to the candidate TD labels collected, using a dedicated tool that we developed. 

\textbf{Step 3: Clean auto-generated data.}
ImDone, an automatic issues generating tool, can automatically generate issues according to code comments \citep{imDone}. When the text of the comment contains "TODO",
imDone automatically generates an issue and tags it as “debt”. Such auto-generated issues with label “debt” are invalid: the ImDone team forked many GitHub repositories to test ImDone, producing a large number of issues with label “debt” in a short time, which damaged the true and real state and life cycle of some issues. In addition, since an issue was automatically tagged with a TD label by imDone, it cannot guarantee that the code comment author considered the “TODO” as TD. Hence, such auto-generated issues with label “debt” by imDone should be filtered out to ensure the authenticity and accuracy of the datasets.

\textbf{Step 4: Clean duplicated TD issues.}
Considering that some issues were submitted multiple times on GitHub, duplicated issues should be cleaned to ensure the uniqueness of data.
Duplicated issues are defined as follows: if the Repo, Reporter, RepTime, IssueTitle, IssueDesc of issues are exactly the same, these issues are considered duplicated. Only one issue will retain in the dataset.


\textbf{Step 5: Clean issues with TD labels without the meaning of TD.}
Some issues tagged with TD labels do not have practical meaning related to TD, but have repository-specific meaning. 
For example, label "TD" does not refer to technical debt in repository \textit{w3c/wot}.
Therefore, such issues should be filtered out.



\begin{table}[]
\centering
\caption{TD labels to be chosen.}
\begin{tabular}{lll}
\hline
\multicolumn{3}{c}{\textbf{TD label}}         \\ \hline
technical debt          & td                        & code debt   \\
technicaldebt           & kind/tech-debt            & test debt  \\
technical-debt          & refactoring / tech debt   & design debt       \\
techdebt                & infrastructure debt       & build debt\\
tech-debt               & requirements debt         & doc debt    \\
tech debt               & architectural debt        & defect debt        \\
debt                    & versioning debt           & ux/ui debt\\ \hline
\end{tabular}
\label{table:labelsetting}
\end{table}

\subsubsection{Data collection for DS2}\label{chap:Data collection for DS2}
In order to explore the understanding of practitioners on the concept of TD, we built dataset DS2 in the following two steps:

\textbf{Step1: Sample TD issues.}
According to Israel's theory of determining sample size \citep{Is1992}, we randomly sampled the resolved TD issues in dataset DS1, setting the margin of error as 5\% and the confidence level as 99\%. The reason for choosing resolved TD issues is that the status of such TD issues is stable.

\textbf{Step2: Analyze the sample manually.}
In this step, we manually tagged the TD type of and the QA affected by each TD issue in DS2 to answer RQ2. The definitions of TD types can be found in \citep{LiAvLi2015}, and the QAs are adopted from the product QAs defined in the ISO 25010 standard \citep{ISO/IEC2011}. In the tagging process, if an issue involves multiple TD types or QAs, we chose the major one. The tagging process is described as follows: 
At first, two researchers made a pilot tagging on 50 TD issues to reach a consensus on the understanding of the TD types and QAs. All disagreements were recorded, and a third researcher joined in to discuss until the three researchers reached a consensus. After this step, the researchers reached an agreement on the scope and boundary of TD types and QAs. 
Later, the two researchers independently tagged a round of 50 TD issues and the Cohen's Kappa coefficient indicating the agreement between the two researchers was calculated \citep{Co1960}. If the coefficient was less than 0.8, the issues with inconsistent tagging results were revisited among the three researchers until agreement was reached. The next round of tagging was performed until the Cohen's Kappa coefficient is greater than 0.8.
After that, the remaining issues were divided into two parts and each was tagged by one of the two researchers independently.

\subsubsection{Data collection for DS3}
In order to understand the characteristics of TD issues, we additionally added a dataset DS3, containing resolved TD issues and resolved non-TD issues. The data collection process for DS3 consists of two steps: 

\textbf{Step1: Select TD issues from specific repositories.} Specifically, this step is to select resolved TD issues of repositories with not less than 100 resolved TD issues from DS1.

\textbf{Step2: Select non-TD issues from specific repositories. }To be specific,  this step is to download and store all the resolved non-TD issues of the repositories mentioned in Step1 from GitHub.

The reason for selecting such repositories is that we believed that the participants of such repositories are more likely to have mature knowledge about TD, and the status of a resolved issue is more stable than an unresolved one.

\subsubsection{Extra data}\label{chart:extra data}
In order to understand the overall issues on GitHub and compare them with TD issues, we collected the number of newly reported issues on GitHub each year, the number of newly reported resolved issues on GitHub each year, the number of repositories newly adopting TDM on GitHub each year, the number of issues of each repository, the number of resolved issues of each repository. 
To understand the maintenance status of each repository, the creating time of the last pull request for each repository was collected as well.
Since the extra data can be retrieved from GitHub by simply using the GitHub APIs, the collection process of extra data is not shown in Figure \ref{fig:datacollectprocess}.

\subsection{Data Analysis}\label{data_analysis}
\subsubsection{Popularity of adopting TDM (RQ1)}
To answer RQ1, we analyze dataset DS1 from three aspects: (1) trend of the number of repositories newly adopting TDM, (2) trend of the number of newly reported TD issues, and (3) distribution of ratio of repositories over the number of TD issues. The time when the first TD issue is reported for a repository is considered as the time of the repository adopting TDM. A repository containing one or more TD issues is considered as it adopting TDM. Every TD issue is intentionally tagged with a TD label by a practitioner, which means that the TD identification activity of TDM is actually performed by the practitioner of the repository. Thus, we consider that the practitioner actively adopts TDM for the repository even it contains only one single TD issue.



Besides descriptive statistics, the compound annual growth rate (CAGR) was used to measure the growth trend of the number of repositories newly adopting TDM and newly reported TD issues in whole DS1. CAGR is defined as: 
\begin{equation}
CAGR=((n_2/n_1)^{1/(y_2-y_1)}-1)\times{100\%}
\end{equation}
where $n_1$ and $n_2$ are the numbers in the years $y_1$ and $y_2$, respectively.

\subsubsection{Understanding on the TD concept (RQ2)}
QAs affected by TD are a central concern in TD research \citep{LiAvLi2015}, and issue labels tagged to TD issues can reflect the characteristics of TD issues to some extent. Thus, we investigate the understanding of practitioners on the TD concept from two perspectives: (1) QAs affected by TD, and (2) co-occurring issue labels with TD labels. 

\textbf{QAs affected by TD.}
As described in Section \ref{chap:Data collection for DS2}, we manually analyzed a sample dataset (\textit{i.e.}, DS2) of all TD issues to explore their TD types and the QAs affected. We then analyzed how each QA is affected by TD. From the QA perspective, we analyzed the TD types affecting each QA, and the extent of the impact. 

\textbf{Co-occurring labels with TD labels.} 
To understand characteristics of TD issues in terms of co-occurring labels, descriptive statistics were used to analyze the number of occurrences of TD labels and the number of labels accompanying TD labels. After obtaining all the issue labels, we further screened all the issue labels to pick out labels with clear meaning and general purposes. The screening process includes three steps: 
a) keep clear and general issues (i.e., not specific to a certain project/repository and not describing the status of the issue), 
b) filter out labels that occur for less than 100 times, and 
c) combine labels with the same meaning. 

\subsubsection{Identification of TD issues (RQ3)} \label{chap:studydesignforTDidentification}
On GitHub, we can trace the timeline of the labeled events (\textit{i.e.}, data item LabeledEvent) pertaining to each issue to further analyze the characteristics of TD issues. A LabeledEvent contains information, such as the person who tagged a label, the label name, the time of tagging the label to the issue. In the following, RQ3 is analyzed from three perspectives. 
    
\textbf{(1) When TD issues are identified.}
We considered the time when a TD label is assigned to the issue as the time when TD is identified for an issue. 
We counted the difference between the reported time and the identification time for each  TD issue. Further, TD identification time can happen at two possible points of time: a) the time when an issue is reported and b) a specific time after an issue is opened to resolve. 


\textbf{(2) Who identifies TD issues.}
To understand the participants' awareness of TDM, we analyzed the participants from several viewpoints. First, we examined the overall distribution and characteristics of participants. Second, we focused on issue reporters who are a key type of participants to identify TD issues. GitHub defines a set of relationships between issue reporter and repository, including: COLLABORATOR, CONTRIBUTOR, FIRST\_TIMER, FIRST\_TIME\_CONTRIBUTOR, MANNEQUIN, MEMBER, OWNER, and NONE. Please refer to Table \ref{table:relationship}
for their detailed definitions. We calculated the distribution of the reporters who identified TD issues. 

\textbf{(3) Phase of adopting TDM in the development lifecycle.} For each repository, let $Period_{TD}$ denote the period between the time when the first TD issue is identified and the time when the first issue is reported, let $Period_{Full}$ denote the period between the time when the last pull request is handled and the time when the first issue is reported, and $Phase_{TD}$ is defined as
\begin{equation}
Phase_{TD}=Period_{TD}/Period_{Full} . 
\end{equation}
Then, $Phase_{TD} \geq 0$, and a smaller $Phase_{TD}$ means earlier awareness of TDM for the repository.


\begin{table*}[]
\centering
\caption{Relationships between issue reporter and repository.}
\scalebox{0.95}{
\begin{tabular}{ll}
\hline
\textbf{Relationship}    & \textbf{Description}                                             \\ \hline
COLLABORATOR             & The issue reporter has been invited to collaborate on the repository.        \\
CONTRIBUTOR              & The issue reporter has previously committed to the repository.               \\
FIRST\_TIMER             & The issue reporter has not previously committed to GitHub.                   \\
FIRST\_TIME\_CONTRIBUTOR & The issue reporter has not previously committed to the repository.           \\
MANNEQUIN                & The issue reporter is a placeholder for an unclaimed user.                   \\
MEMBER                   & The issue reporter is a member of the organization that owns the repository. \\
OWNER                    & The issue reporter is the owner of the repository.                           \\
NONE                     & The issue reporter has no association with the repository.                   \\ \hline
\end{tabular}}
\label{table:relationship}
\centering
\end{table*}

\subsubsection{Resolution of TD issues (RQ4)}
To answer RQ4, we made analysis in four aspects: 

\textbf{(1) Proportion of resolved TD issues.}
For each repository in dataset DS1, we calculated the proportion of resolved TD issues over the total TD issues based on data item IsResolved (\textit{i.e.}, D11). We then calculated the distribution of the number of repositories over the proportion of resolved TD issues. 
In addition, we also calculated the proportion of TD issues resolved and the proportion of resolved issues on the whole GitHub from 2009 to 2020.

 
\textbf{(2) Open time of resolved TD issues.}
At first, we calculated the open time of each resolved TD issue as the difference between ResTime and RepTime of the issue. In the subsequent data analysis, we calculated the distribution of the proportion of resolved TD issues to all resolved TD issues in DS1 over the open time, and observed the characteristics of resolved TD issues in terms of open time; we also compared the differences in the distribution of resolved TD issues and resolved non-TD issues in DS3.

\textbf{(3) Characteristics of resolved TD issues.}


We explored the TD issue resolution process based on the characteristics of resolved issues. Only resolved issues were used as the data source because their status is much more stable than unresolved issues.

On GitHub, each issue allows participants to discuss the issue by posting comments on it, in addition to the title and descriptive text provided by the reporter. When a participant posts a comment on the issue, a comment event (\textit{i.e.}, data item CmntEvent) for the issue is generated.

Then, we investigated whether the resolved TD issue and the resolved non-TD issue differ significantly in terms of the issue content and the corresponding comment events, including four characteristics: the size of the title and description (IssueSize), the number of all CmntEvents (CmntNo), the size of the content of all CmntEvents (CmntSize), the number of the participants who added all CmntEvents (ParticipantNo). 
We ran Mann-Whitney U Tests \citep{Fi2013} on the four issue characteristics of resolved TD issues and resolved non-TD issues in DS3, and corrected the \textit{p-value} of Mann-Whitney U Tests by the Bonferroni correction in account for the multiple comparisons problem \citep{du1961}.



\textbf{(4) Reopen of resolved TD issues.}
On GitHub, if an issue is resolved, the status of the issue is updated to "closed". However, an issue can be changed to "opened" again, which means that the issue may need further discussion and fixes. We considered that the probability for a resolved issue to reopen reflects the quality of the previous fix of the issue. Thus, we calculated the probabilities of resolved TD issues and non-TD issues based on the reopened events (\textit{i.e.}, data item ReopenedEvent) of all issues in DS3.

\subsubsection{Continuity of TDM for the repositories with TD labels (RQ5)}

To better understand the continuity of TDM of repositories with TD labels, we defined three levels of TDM continuity (namely Abandoned, Unclear, and Consistent) for such repositories. 
First, for convenient description, we define several notations.
$Period_{PT}$ denotes the period between the time when the last pull request is handled and the time when the last TD issue is reported.   
$Period_{TT}$ denotes the period between the time when the last TD issue is reported and the time when the first TD issue is reported. 
$Period_{Y}$ is a constant, and defined as: suppose that the sequence of time differences between the tagging time of each pair of adjacent TD issues of each repository is known, and $Period_{Y}$ is the 99th percentile of this sequence of time differences.  
$Count_{R}$ denotes the number of TD issues of the repository. 
$Count_{Z}$ is a constant, which denotes the 75th percentile of the number of TD issues of the repositories in DS1. 
Then, the conditions for the three TDM continuity levels for a repository are defined as follows.\\
\indent 1) \textbf{Abandoned}: the repository has abandoned TDM, \textit{i.e.}, $Period_{PT} > Period_{Y}$. \\  
\indent 2) \textbf{Unclear}: the repository does not explicitly show TDM is adopted continuously, \textit{i.e.},  
($Period_{PT}\leq Period_{Y}$) $\land$ (($Period_{TT} \leq Period_{Y}$)$\vee$($Count_{R} < Count_{Z}$)) is true.\\
\indent 3) \textbf{Consistent}: the repository keeps adopting TDM, \textit{i.e.}, ($Period_{PT}\leq{Period_{Y}}$)$\land$($Period_{TT}> Period_{Y}$)$\land$($Count_{R}\geq{Count_{Z}}$) is true. 


Subsequently, we analyzed the levels of all repositories in DS1, and further analyzed the differences between repositories at level Abandoned and level Consistent. We focused on the differences between the two levels of repositories in terms of total issues and TD issues. Seven characteristics for each repository were extracted: the number of total issues (RepoIssueNo), the proportion of total issues resolved (RepoResolvedIssuePro),  the number of TD issue reporters (RepoTDRepoterNo), the number of TD issues (RepoTDIssueNo),  the proportion of TD issues resolved (RepoResolvedTDIssuePro), the average open time of TD issues (RepoOpentimeAvg),  and the average length of the  comments of TD issues (RepoCmntSizeAvg).

Furthermore, we used Mann-Whitney U Tests \citep{Fi2013} on the characteristics to compare if there is a significant difference between the repositories of level Abandoned and those of level Consistent. In addition, given the problem of multiple comparisons, we performed the Bonferroni correction to correct the \textit{p-value} of Mann-Whitney U tests \citep{du1961}.

\section{Results}\label{chap:results}

\subsection{Overview of the Obtained Datasets}\label{OverviewOfDataSet}
Following the data collection procedure described in Section \ref{DataCollection}, we obtained DS1, DS2, and DS3.
48,915 issues were retrieved by the searches with candidate TD labels, and finally \textbf{DS1} includes 35,278 TD issues collected from 3,598 repositories after data cleaning steps. 
Noticeably, among the 35,278 TD issues, there are 914 issues whose LabeledEvents cannot be retrieved from GitHub, but their TD labels can be retrieved. Considering the relatively large number of such issues, we kept them in the final dataset (DS1) of TD issues. This may influence the results related to the LabeledEvents of TD issues. Specifically, when analyzing TD issues related to the LabeledEvents, we used the left 34,364 TD issues excluding these 914 TD issues. 
\textbf{DS2} contains 652 issues randomly extracted from DS1 according to the sampling requirements stated in Section \ref{chap:Data collection for DS2}.
\textbf{DS3} contains 5,202 resolved TD issues and 205,927 resolved non-TD issues of the 24 repositories with at least 100 resolved TD issues. The list of these repositories is available online \citep{Repo&NoresolvedTDissue}. All the three datasets (\textit{i.e.}, DS1, DS2, and DS3) are also available online \cite{AllDatasets2022}.

\subsection{RQ1: Popularity of adopting TDM}\label{RQ1Results}
We studied the popularity of adopting TDM in OSS development practice in the following three aspects.

\subsubsection{Trend of the number of repositories newly adopting TDM}
\begin{figure}[htpb]
    \centering
    \includegraphics[width=8.0cm]{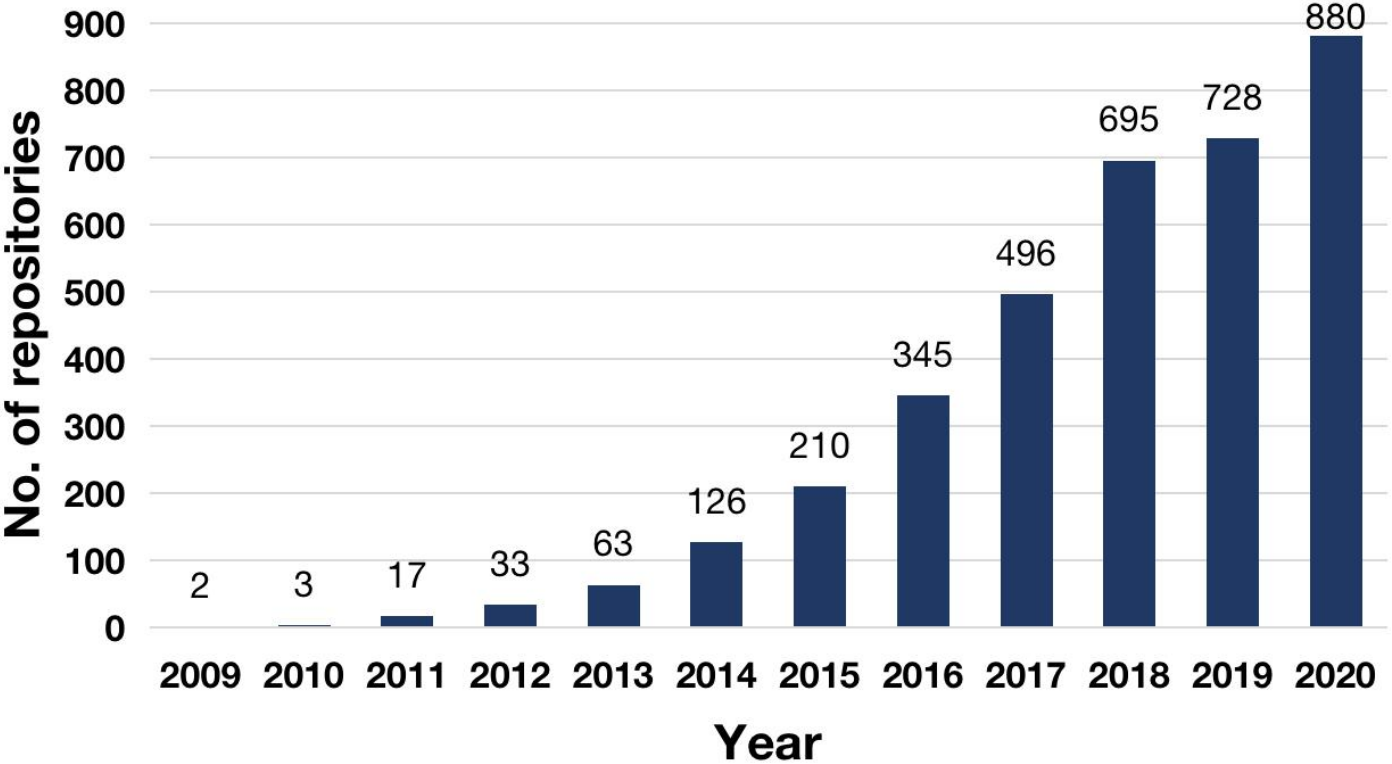}
    \caption{Distribution of the number of repositories newly adopting the TDM practice over years (RQ1).}
    \label{fig:rq1_RepoNo}
\end{figure}

 \textbf{Figure \ref{fig:rq1_RepoNo} shows that the number of repositories newly adopting TDM had been lastingly increasing over the years from 2 in 2009 to 880 in 2020, and the CAGR is 73.9\%.} In the same period, the number of newly created repositories had also been continuously growing from 80,318 in 2009 to 23,807,596 in 2020, and its CAGR is 67.8\%, which is little lower than the CAGR of the number of repositories newly adopting TDM.


\subsubsection{Trend of the number of newly reported TD issues}
\begin{figure}[htpb]
    \centering
    \includegraphics[width=8.0cm]{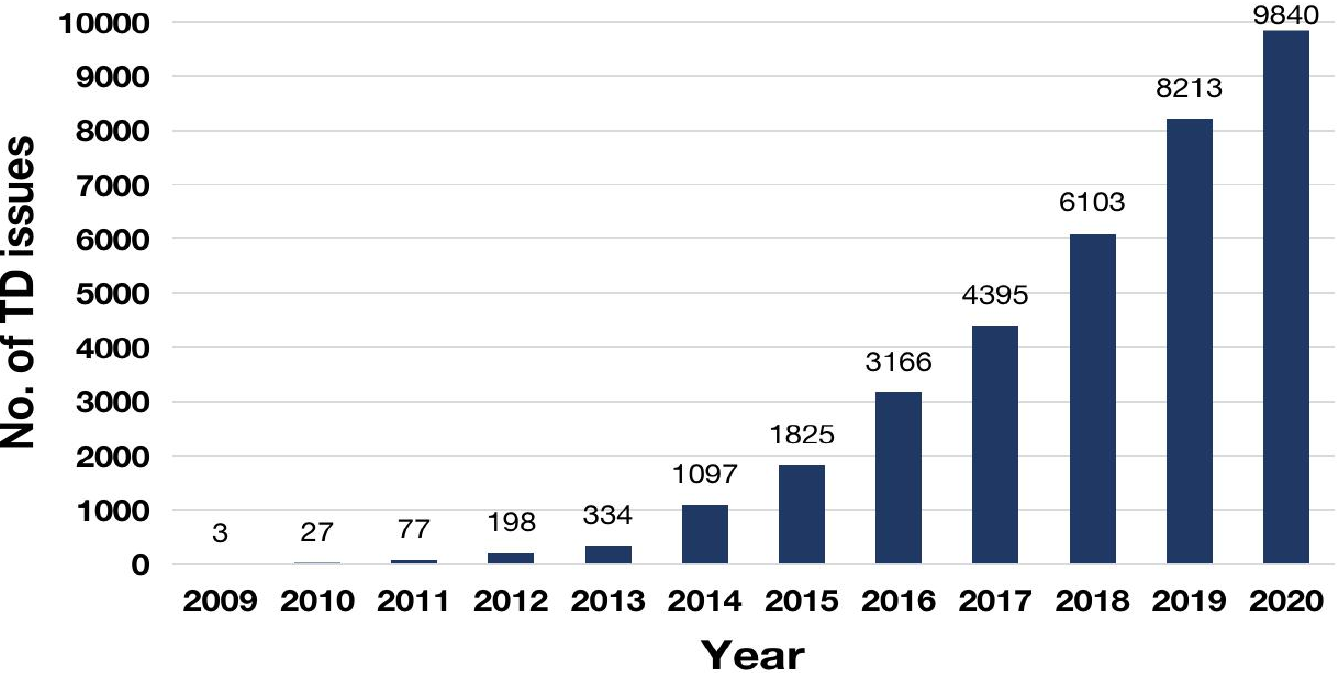}
    \caption{Distribution of the number of newly reported TD issues over years (RQ1).}
    \label{fig:rq1_tdissues}
\end{figure}

Figure \ref{fig:rq1_tdissues} shows that \textbf{the number of newly reported TD issues had been continuously increasing over the years from 3 in 2009 to 9,840 in 2020, and the CAGR is 108.7\%.} Meanwhile, the number of newly reported issues had been continuously growing from 92,552 in 2009 to 58,335,696 in 2020, and its CGAR is 79.7\%, which is much lower than the CGAR of the number of newly reported TD issues.
In addition, the CGAR of the proportion of newly reported TD issues over the total newly reported issues was also calculated, and the CGAR is 16.2\%. However, the proportion of newly reported non-TD issues over the total newly reported issues decreases with a CGAR close to 0.0\% (i.e., -0.00124\%) during the 12 years. 

\subsubsection{Distribution of the ratio of repositories over the number of TD issues}
\begin{figure}[htpb]
    \centering
    \includegraphics[width=8cm]{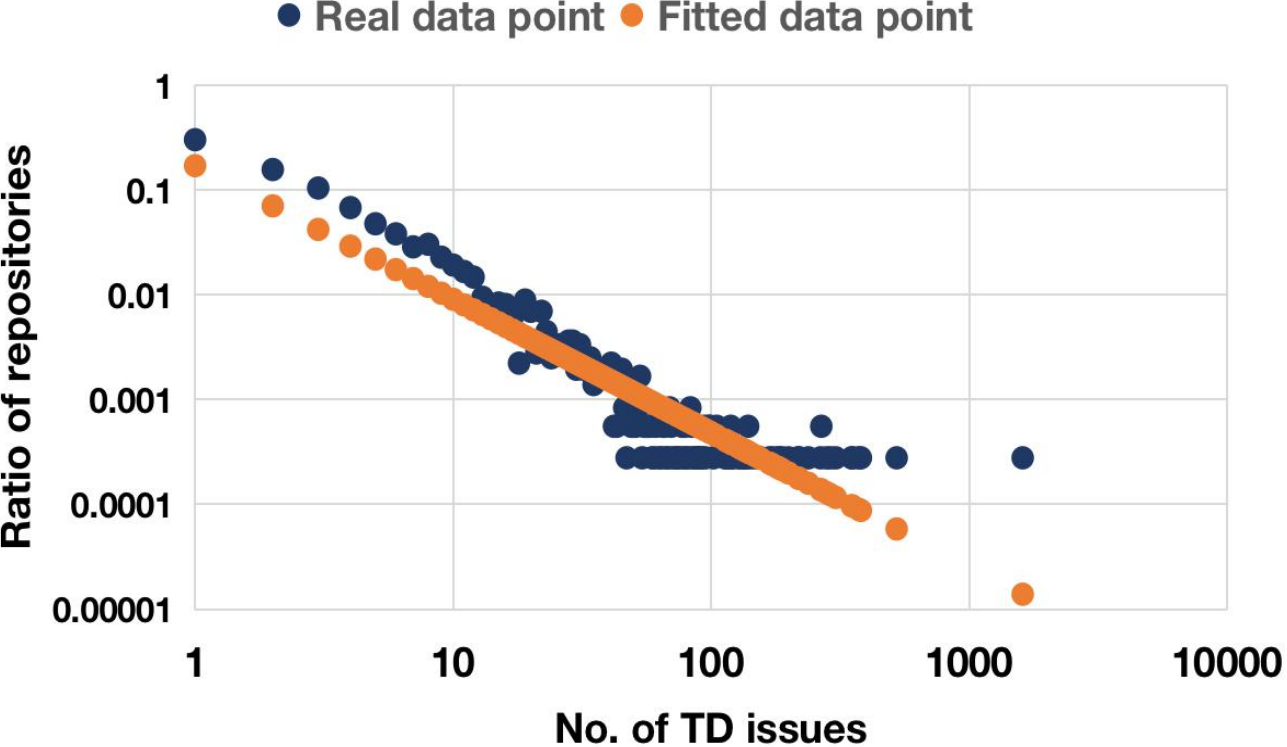}
    \caption{Distribution of ratio of repositories with TD labels over the number of TD issues (RQ1).}
    \label{fig:rq1_RepoDistribution}
\end{figure}

Among the total 3,598 repositories with TD labels, there are 1,084 (30.1\%), 563 (15.6\%), and 2,937 (81.6\%) repositories that each has only one, two, and less than ten TD issues, respectively. Figure \ref{fig:rq1_RepoDistribution} depicts the distribution of the ratio of repositories over the number of TD issues. 
This distribution obeys a power law \citep{BaAl1999}, \textit{i.e.}, $y=bx^{-a}$, where $y$ denotes the ratio of repositories, $x$ denotes the number of TD issues, $a$ and $b$ are constants. The fitted results are that $a=1.275$ and $b=0.170$, \textit{i.e.}, $y=0.170x^{-1.275}$ (see Figure \ref{fig:rq1_RepoDistribution}).



\begin{center}
\fbox{\parbox{0.475\textwidth}{\textbf{Summary (RQ1)}: The number of repositories adopting TDM and the number of newly reported TD issues had been continuously and rapidly increasing, and their CAGRs are 73.9\% and 108.7\%, respectively. The distribution of the ratio of repositories over the number of TD issues obeys a power law. }}
\end{center}


\subsection{RQ2: Understanding on the TD Concept}\label{RQ2Results}


\subsubsection{QAs affected by TD} 
During our manual analysis, we found that the existing widely-used TD classification proposed in \citep{LiAvLi2015} cannot fully cover all the TD issues. Therefore, except for the ten TD types proposed in \citep{LiAvLi2015}, we added “Deployment TD” as a new TD type. 
\textbf{Deployment TD} refers to flaws that slow down software deployment or make software deployment unnecessarily complicated. For example, from issue \#151\footnote{https://github.com/konveyor/pelorus/issues/151} of project \textit{konveyor/pelorus}, we can see that some warnings have affected the deployment of Kubernetes, 
 and thus this issue contains deployment TD. 
 
Table \ref{table:qa} shows the results of our manual analysis. For each QA, we calculated the number and percentage of which TD types it was affected by. In addition, for each column, we fill the cells in green according to the value of the number, in order to show more clearly the relationship between affected QAs and TD types. The darker the color of a cell, the larger the number in the cell. As shown in this table, \textbf{the top three QAs most affected by TD issues are maintainability, reliability, and functional suitability, and the top three TD types most likely contained in TD issues are design TD, architectural TD, and test TD}.
 

In the following, we present the results on how each QA is affected by TD. 
(1) \textbf{Maintainability} is mainly affected by design TD, architectural TD, and test TD, accounting for 30.9\%, 27.3\%, and 23.8\%, respectively. 
(2) \textbf{Reliability} is mainly affected by five types of TD, of which defect TD has the greatest impact on this QA, accounting for 34.0\%. 
(3) For \textbf{functional suitability}, design TD is the most influential TD type, followed by architectural TD. 
(4) For \textbf{usability}, design TD, architectural TD, and requirements TD are the top three TD types that account for more than 10.0\%. Among the three types of TD, design TD with 58.8\% ranks in the first place. 
(5) \textbf{Performance efficiency} is most affected by design TD and architectural TD, accounting for 51.5\% and 33.3\%, respectively. 
(6) \textbf{Portability} is mainly affected by architectural TD, accounting for more than a half. 
Portability requires a developer to consider the relationship between the software system and the computing platforms during design. Thus, architectural TD has a great impact on portability. 
(7) \textbf{Security} is mainly affected by design TD, architectural TD, and test TD, accounting for 23.1\%, 46.2\%, and 15.4\%, respectively. 
(8) \textbf{Compatibility} is mainly affected by infrastructure TD with 36.4\%.

\begin{table*}[]
\centering
\caption{How each QA is affected by TD (RQ2).}
\scalebox{0.70}{
\begin{tabular}{|l|l|l|l|l|l|l|l|l|l|}
\hline
\rowcolor[HTML]{BDD7EE} 
& {\color[HTML]{333333} \textbf{Maintainability}} & {\color[HTML]{333333} \textbf{Reliability}} & \textbf{Functional suitability}    & \textbf{Usability}                 & \textbf{Performance efficiency}    & \textbf{Portability}               & \textbf{Security}      & \textbf{Compatibility}   & \textbf{All QAs}         \\ \hline
\cellcolor[HTML]{BDD7EE}\textbf{Design TD}         & \cellcolor[HTML]{ACDCBA}122, 30.9\%             & \cellcolor[HTML]{C5E6D0}20, 21.3\%          & \cellcolor[HTML]{91D1A3}21, 41.2\% & \cellcolor[HTML]{63BE7B}20, 58.8\% & \cellcolor[HTML]{77C68C}17, 51.5\% & \cellcolor[HTML]{D6EDDE}3, 14.3\%  & \cellcolor[HTML]{BFE3CA}3, 23.1\% & \cellcolor[HTML]{CCE9D5}2, 18.2\%   & \cellcolor[HTML]{63BE7B}208, 31.9\%\\ \hline
\cellcolor[HTML]{BDD7EE}\textbf{Architectural TD}  & \cellcolor[HTML]{B5E0C2}108, 27.3\%             & \cellcolor[HTML]{D0EBD9}16, 17.0\%          & \cellcolor[HTML]{BAE2C6}13, 25.5\% & \cellcolor[HTML]{DEF0E5}4, 11.8\%  & \cellcolor[HTML]{A6D9B5}11, 33.3\% & \cellcolor[HTML]{63BE7B}12, 57.1\% & \cellcolor[HTML]{81CA95}6, 46.2\% & \cellcolor[HTML]{CCE9D5}2, 18.2\%  & \cellcolor[HTML]{7EC992}172, 26.4\%\\ \hline
\cellcolor[HTML]{BDD7EE}\textbf{Test TD}           & \cellcolor[HTML]{BFE3CA}94, 23.8\%              & \cellcolor[HTML]{DBEFE3}12, 12.8\%          & \cellcolor[HTML]{EDF6F2}3, 5.9\%   & \cellcolor[HTML]{FCFCFF}0, 0.0\%   & \cellcolor[HTML]{FCFCFF}0, 0.0\%   & \cellcolor[HTML]{FCFCFF}0, 0.0\%   & \cellcolor[HTML]{D3ECDC}2, 15.4\% & \cellcolor[HTML]{E4F3EA}1, 9.1\%  & \cellcolor[HTML]{AADBB9}112, 17.2\%\\ \hline
\cellcolor[HTML]{BDD7EE}\textbf{Defect TD}         & \cellcolor[HTML]{FCFCFF}0, 0.0\%                & \cellcolor[HTML]{A4D9B3}32, 34.0\%          & \cellcolor[HTML]{E3F2E9}5, 9.8\%   & \cellcolor[HTML]{F5F9F9}1, 2.9\%   & \cellcolor[HTML]{FCFCFF}0, 0.0\%   & \cellcolor[HTML]{F0F7F4}1, 4.8\%   & \cellcolor[HTML]{E8F4EE}1, 7.7\%  & \cellcolor[HTML]{E4F3EA}1, 9.1\%  & \cellcolor[HTML]{DFF1E6}41, 6.3\% \\ \hline
\cellcolor[HTML]{BDD7EE}\textbf{Code TD}           & \cellcolor[HTML]{EFF7F4}20, 5.1\%               & \cellcolor[HTML]{D9EEE0}13, 13.8\%          & \cellcolor[HTML]{EDF6F2}3, 5.9\%   & \cellcolor[HTML]{EDF6F2}2, 5.9\%   & \cellcolor[HTML]{F5F9F9}1, 3.0\%   & \cellcolor[HTML]{FCFCFF}0, 0.0\%   & \cellcolor[HTML]{FCFCFF}0, 0.0\%  & \cellcolor[HTML]{FCFCFF}0, 0.0\%  & \cellcolor[HTML]{E0F1E7}39, 6.0\%\\ \hline
\cellcolor[HTML]{BDD7EE}\textbf{Infrastructure TD} & \cellcolor[HTML]{EDF6F2}23, 5.8\%               & \cellcolor[HTML]{FCFCFF}0, 0.0\%            & \cellcolor[HTML]{F7FAFB}1, 2.0\%   & \cellcolor[HTML]{FCFCFF}0, 0.0\%   & \cellcolor[HTML]{F5F9F9}1, 3.0\%   & \cellcolor[HTML]{CAE8D4}4, 19.0\%  & \cellcolor[HTML]{E8F4EE}1, 7.7\%  & \cellcolor[HTML]{9BD5AB}4, 36.4\% & \cellcolor[HTML]{E4F3EA}34, 5.2\%\\ \hline
\rowcolor[HTML]{FCFCFF} 
\cellcolor[HTML]{BDD7EE}\textbf{Documentation TD}  & \cellcolor[HTML]{F3F8F7}15, 3.8\%               & 0, 0.0\%                                    & \cellcolor[HTML]{F7FAFB}1, 2.0\%   & \cellcolor[HTML]{E6F3EC}3, 8.8\%   & 0, 0.0\%                           & 0, 0.0\%                           & 0, 0.0\%                          & 0, 0.0\%               & \cellcolor[HTML]{EFF7F4}19, 2.9\% \\ \hline
\cellcolor[HTML]{BDD7EE}\textbf{Build TD}          & \cellcolor[HTML]{F7FAFA}9, 2.3\%                & \cellcolor[HTML]{FAFBFD}1, 1.1\%            & \cellcolor[HTML]{FCFCFF}0, 0.0\%   & \cellcolor[HTML]{FCFCFF}0, 0.0\%   & \cellcolor[HTML]{E5F3EB}3, 9.1\%   & \cellcolor[HTML]{FCFCFF}0, 0.0\%   & \cellcolor[HTML]{FCFCFF}0, 0.0\%  & \cellcolor[HTML]{E4F3EA}1, 9.1\%  & \cellcolor[HTML]{F3F9F7}14, 2.1\%\\ \hline
\rowcolor[HTML]{FCFCFF} 
\cellcolor[HTML]{BDD7EE}\textbf{Requirements TD}   & 0, 0.0\%                                        & 0, 0.0\%                                    & \cellcolor[HTML]{E8F4EE}4, 7.8\%   & \cellcolor[HTML]{DEF0E5}4, 11.8\%  & 0, 0.0\%                           & 0, 0.0\%                           & 0, 0.0\%                          & 0, 0.0\%           
& \cellcolor[HTML]{F7FAFB}8, 1.2\% \\ \hline
\rowcolor[HTML]{FCFCFF} 
\cellcolor[HTML]{BDD7EE}\textbf{Versioning TD}     & \cellcolor[HTML]{FBFCFE}3, 0.8\%                & 0, 0.0\%                                    & 0, 0.0\%                           & 0, 0.0\%                           & 0, 0.0\%                           & \cellcolor[HTML]{F0F7F4}1, 4.8\%   & 0, 0.0\%                          & 0, 0.0\%               & \cellcolor[HTML]{FAFCFE}4, 0.6\%           \\ \hline
\rowcolor[HTML]{FCFCFF} 
\cellcolor[HTML]{BDD7EE}\textbf{Deployment TD}     & 1, 0.3\%                                        & 0, 0.0\%                                    & 0, 0.0\%                           & 0, 0.0\%                           & 0, 0.0\%                           & 0, 0.0\%                           & 0, 0.0\%                          & 0, 0.0\%              & \cellcolor[HTML]{FCFCFF}1, 0.2\%          \\ \hline
\rowcolor[HTML]{FCFCFF} 
\cellcolor[HTML]{BDD7EE}\textbf{Total}     &\cellcolor[HTML]{63BE7B}395, 100.0\%         &\cellcolor[HTML]{63BE7B}94, 100.0\%      &\cellcolor[HTML]{63BE7B}51, 100.0\%    &\cellcolor[HTML]{63BE7B} 34, 100.0\%          &\cellcolor[HTML]{63BE7B}33, 100.0\%       &\cellcolor[HTML]{63BE7B}21, 100.0\%     &\cellcolor[HTML]{63BE7B}13, 100.0\%                          &\cellcolor[HTML]{63BE7B}11, 100.0\%       & \cellcolor[HTML]{63BE7B}652, 100.0\%                   \\ \hline
\end{tabular}
}
\label{table:qa}
\end{table*}

\subsubsection{Co-occurring labels with TD labels}
Table \ref{table:TDLabels} lists the TD labels used in the repositories on GitHub and the number of their occurrences. Fifteen TD labels are used, \textit{technical debt} was used most frequently by 11,787 issues, and \textit{doc debt} was used most scarcely by only 12 issues. 
\textbf{There are 3,974 co-occurring labels and most of them occurred just for a few times.} After filtering, the final co-occurring labels and their numbers of the occurrences are shown in Table \ref{table:co-occurringlabels}. 
As we can see, \textbf{the most frequent co-occurring label is \textit{enhancement/improvement}, followed by \textit{bug}.
Co-occurring labels may indicate specific characteristics of TD issues.} For instance, some co-occurring labels are associated with the purposes (\textit{e.g.}, \textit{enhancement}) and expectations (\textit{e.g.}, \textit{nicetohave}) of the TD issues.
Some co-occurring labels (\textit{e.g.}, \textit{testing}) indicate the TD types contained by the issues, while some others (\textit{e.g.}, \textit{security}) indicate the QAs affected by TD.

\begin{table}[]
\caption{TD labels and their occurrences (RQ2).}
\centering
\scalebox{1.0}{
\begin{tabular}{lrlr}
\hline
\textbf{TD label}       & \textbf{\#}    &  \textbf{TD label}       & \textbf{\#}       \\ \hline
technical debt          & 11,787          &  technicaldebt           & 137               \\
tech debt               & 9,341           &  code debt               & 129               \\   
debt                    & 4,294           &  ux/ui debt              & 112               \\
tech-debt               & 3,535           &  design debt             & 37                \\
techdebt                & 2,855           &  refactoring / tech debt & 37                \\
technical-debt          & 2,542           &  td                      & 14                \\
test debt               & 241            &  doc debt                & 12                \\
kind/tech-debt          & 205           \\ \hline
\end{tabular}}
\label{table:TDLabels}
\end{table}

\begin{table}[]
\caption{Labels co-occurring with TD labels (RQ2).}
\centering
\scalebox{0.87}{
\begin{tabular}{lrlr}
\hline
\textbf{Label}           & \textbf{\#}             & \textbf{Label}          & \textbf{\#}     \\ \hline
enhancement, improvement & 2,152                    & task                     & 178   \\      
bug                      & 1,499                    & epic                     & 174   \\
help wanted              & 860                     & code                     & 173   \\
testing, tests, test     & 831                     & should                   & 162   \\
good first issue         & 827                     & backlog                  & 158   \\
refactoring, refactor    & 368                     & security                 & 157   \\
feature, feature request & 348                     & performance              & 157   \\
documentation            & 341                     & area/integration tests   & 148   \\
stale, status/stale      & 318                     & server                   & 145   \\
api                      & 251                     & devops                   & 142   \\
nicetohave               & 233                     & area/unit tests          & 130   \\
must                     & 192                     & infrastructure           & 120   \\
question                 & 191                     & debug                    & 108   \\
wontfix                  & 190                      \\ \hline
\end{tabular}}
\label{table:co-occurringlabels}
\end{table}

\begin{center}
\fbox{\parbox{0.475\textwidth}{\textbf{Summary (RQ2)}: Maintainability, reliability, and functional suitability are the top three QAs affected most by TD issues. Labels \textit{enhancement/improvement} and \textit{bug} are more likely associated with TD labels. }}
\end{center}



\subsection{RQ3: Identification of TD issues}\label{RQ3Results}
We looked into the identification of TD issues from the following three perspectives: when TD issues are identified, who identifies TD issues, and phase of adopting TDM during the development lifecycle. As mentioned in Section \ref{OverviewOfDataSet}, when we needed to analyze the labeled events of the TD issues in DS1, we excluded the 914 TD issues whose labeled events of TD labels were missing. Therefore, we analyzed the left 34,364 TD issues to answer RQ3 from the first two perspectives, and analyzed the full DS1 to answer RQ3 from the third perspective.

\subsubsection{When TD issues are identified}
In dataset DS1, there are 34,480 TD labeled events that record the assignment of TD labels to 34,364 TD issues. 
\textbf{Among the 34,364 TD issues, 16,967 (49.4\%) were tagged with TD labels when they were reported, and the other 17,397 (50.6\%) were tagged with TD labels during the issue resolution process.}
For the 17,397 TD issues, the maximum duration between the issue reporting time and the TD tagging time is 2,784 days, the minimum is 1 day, and the average is 63 days. In addition, 
the 70th percentile is 15 days, and the median is 1 day. 





\subsubsection{Who identifies TD issues}
In dataset DS1, \textbf{all TD issues were identified by 4,526 participants, 1,408 (31.1\%) of which identified only one TD issue}, and 23 (0.5\%) of which identified more than 100 TD issues. On average, each participant identified 7.6 TD issues.
In addition, \textbf{there were 2,438 (69.0\%) repositories in which all TD issues were identified by only one participant}. In more than 3,306 (91.9\%) of all repositories had no more than 4 participants who identified TD issues. 

Among the 34,364 TD issues, \textbf{27,690 (80.5\%) were identified as TD issues by their reporters}, and 6,674 (19.5\%) were identified as TD by other participants other than the issue reporters. 
Through analyzing the 27,690 TD issues, 
we found that the proportion of CONTRIBUTOR and MEMBER among the reporters of the TD issues is relatively high. Table \ref{table:num and pro of relationship} shows the number and proportion of each relationship. Among them, the numbers of FIRST\_TIMER, FIRST\_TIME\_CONTRIBUTOR, and MANNEQUIN are 0, and thus these three relationships do not appear in Table \ref{table:num and pro of relationship}. 





\begin{table}[]
\centering
\caption{The number and proportion of each relationship between issue reporters and repositories of TD issues (RQ3).}
\scalebox{1.00}{
\begin{tabular}{lrr}
\hline
\textbf{Relationship} & \textbf{\#Reporter}                        & \textbf{\%Reporter} \\ \hline
COLLABORATOR & 3,946                                            & 14.3\%                                              \\
CONTRIBUTOR  & 9,236                                            & 33.4\%                                              \\
MEMBER       & 9,052                                            & 32.7\%                                              \\
OWNER        & 4,934                                            & 17.8\%                                              \\
NONE         & 522                                             & 1.9\%                                              \\
Total        & 27,690                                           & 100.0\%       \\\hline                                      
\end{tabular}}
\label{table:num and pro of relationship}
\end{table}

 


\subsubsection{Phase of adopting TDM during the development lifecycle}

\begin{figure}[htpb]
    \centering
    \includegraphics[width=7.0cm]{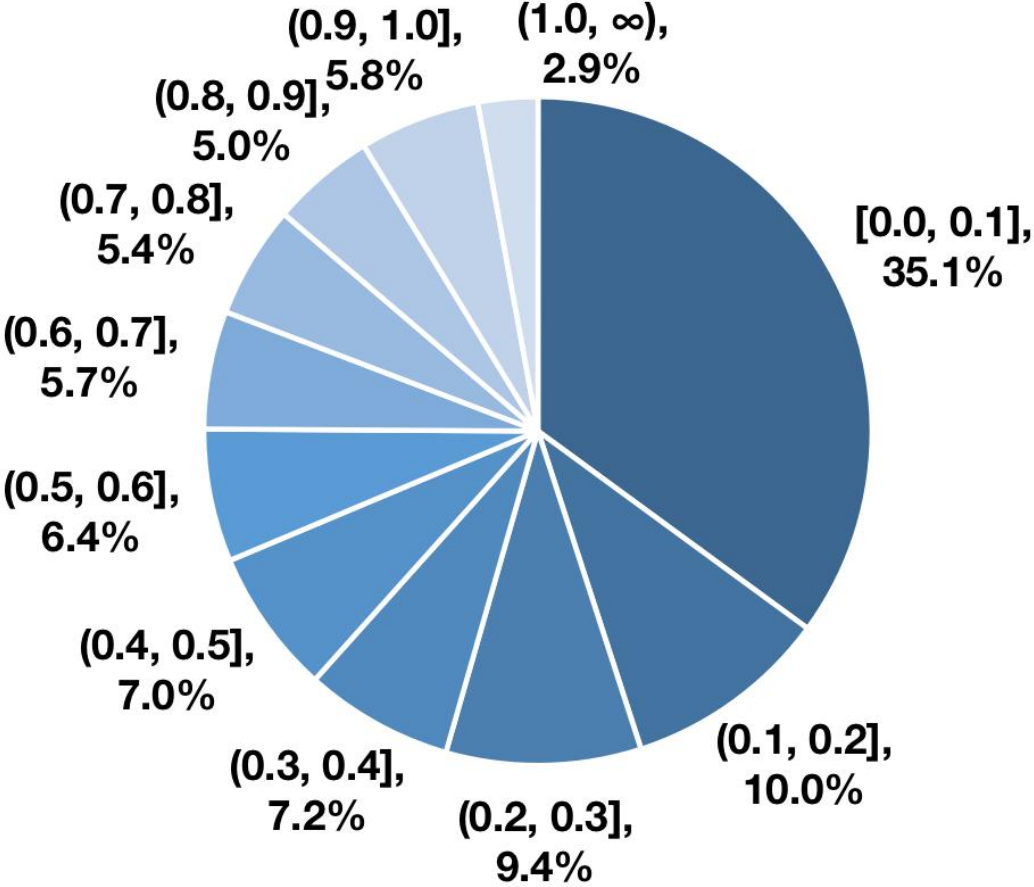}
    \caption{Distribution of percentage of repositories with TD labels over the intervals of $Phase_{TD}$ (RQ3).}
    \label{fig:rq3_RepoDistribution}
\end{figure}

As shown in Figure \ref{fig:rq3_RepoDistribution}, the number of repositories with TD labels tends to decrease as $Phase_{TD}$ increases. Notably, 35.1\% of the repositories have a $Phase_{TD}$ value no more than 0.1, and more than half of the repositories have a $Phase_{TD}$ value less than or equal to 0.3. This means that the participants start investing effort to TDM in the early phase of the development lifecycle.

\begin{center}
\fbox{\parbox{0.475\textwidth}{\textbf{Summary (RQ3)}: Around a half (49.4\%) of TD issues were identified when they were reported, and the others were identified during the resolution process. 31.1\% of the participants identified only one TD issue, and 69.0\% of the repositories were with all TD issues identified by only one participant. 
More than a half of the repositories 
adopt TDM in the early phase of the development lifecycle. }}
\end{center}


\subsection{RQ4: Resolution of TD Issues}\label{RQ4Results}

\subsubsection{Proportion of resolved TD issues}
\begin{figure}[htpb]
    \centering
    \includegraphics[width=8.0cm]{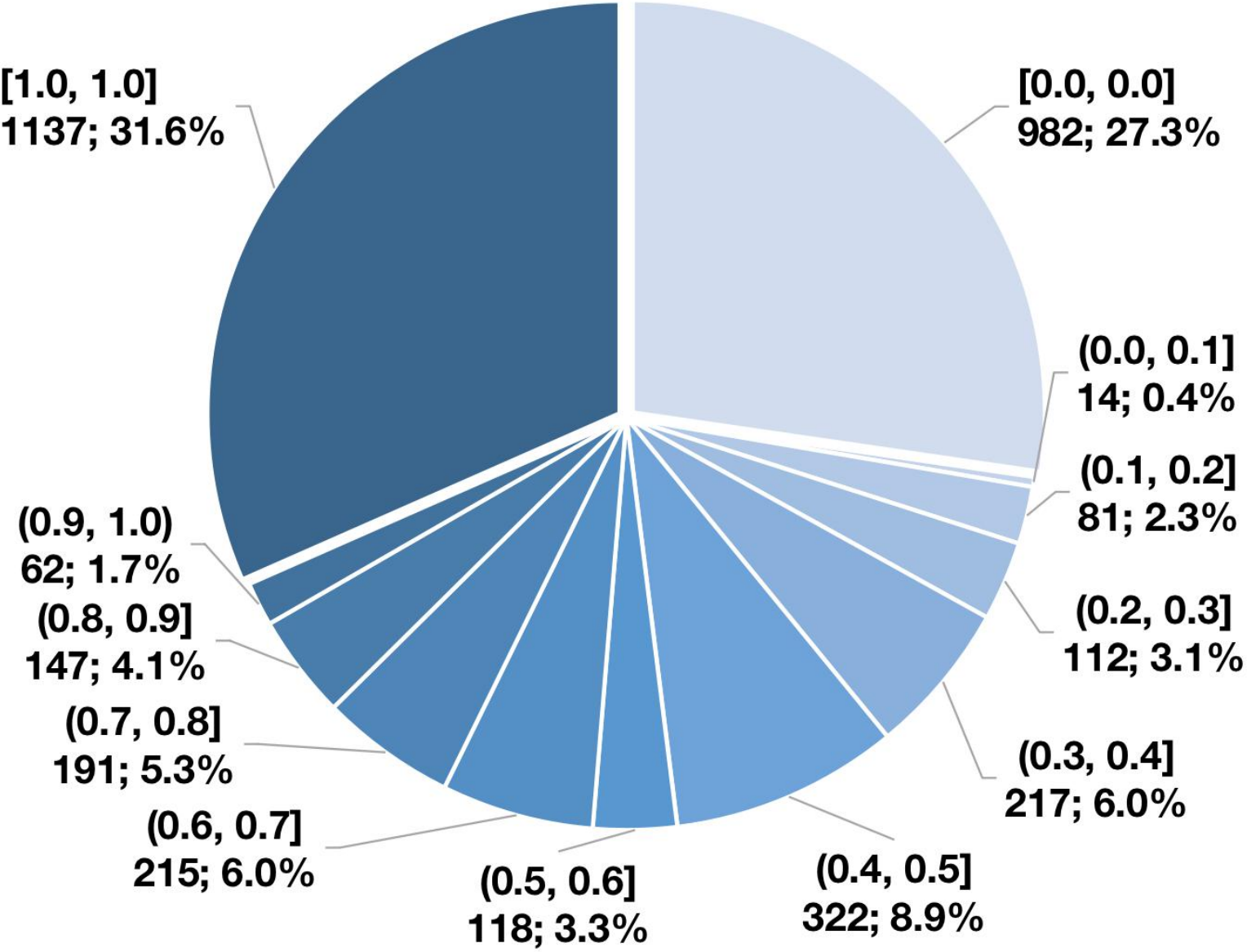}
    \caption{Distribution of the number of repositories over the proportion of resolved TD issues (RQ4).}
    \label{fig:RQ4_dist_repo_prop2}
\end{figure}

The distribution of the number of repositories over the proportion of resolved TD issues is shown in Figure \ref{fig:RQ4_dist_repo_prop2}. The proportion of resolved TD issues is divided into multiple intervals, \textit{e.g.}, [0.0, 0.0], (0.0, 0.1], and [1.0, 1.0]. The number and proportion of repositories in each interval are shown in Figure \ref{fig:RQ4_dist_repo_prop2}. 
\textbf{All TD issues were resolved in 1,137 out of 3,598 (31.6\%) repositories} (corresponding to interval [1.0, 1.0]), and each repository contains 4.5 TD issues on average. 
\textbf{None of the TD issues was resolved in 982 out of 3,598 (27.3\%) repositories} (corresponding to interval [0.0, 0.0]), and  each repository contains 2.1 TD issues on average.  
In each of the rest 1,479 out of 3,598 (41.1\%) repositories, the proportion of resolved TD issues falls into (0.0, 1.0), \textit{i.e.}, part of the TD issues were resolved. 

The median of the proportion of resolved TD issues for all 3,598 repositories in DS1 is 0.582. The total number of all resolved TD issues between 2009 and 2020 is 22,841, the total number of all reported TD issues is 35,278, and thus the proportion of resolved TD issues is 64.7\%. In contrast, the total number of resolved issues and the total number of issues by the end of 2020 are 49,694,427 and 72,813,679, respectively, and hence the proportion of resolved issues is 68.2\%, which is slightly greater than the proportion of resolved TD issues.
Repository \textit{microsoft/vscode} has 1,610 TD issues, and is the only repository that has more than 1,000 TD issues. The proportion of resolved TD issues in this repository is 92.7\%, which is a very high proportion, especially considering that some unresolved TD issues reported recently would be resolved afterwards.

 






\subsubsection{Open time of TD issues}
\begin{figure}[htpb]
    \centering
    \includegraphics[width=8.0cm]{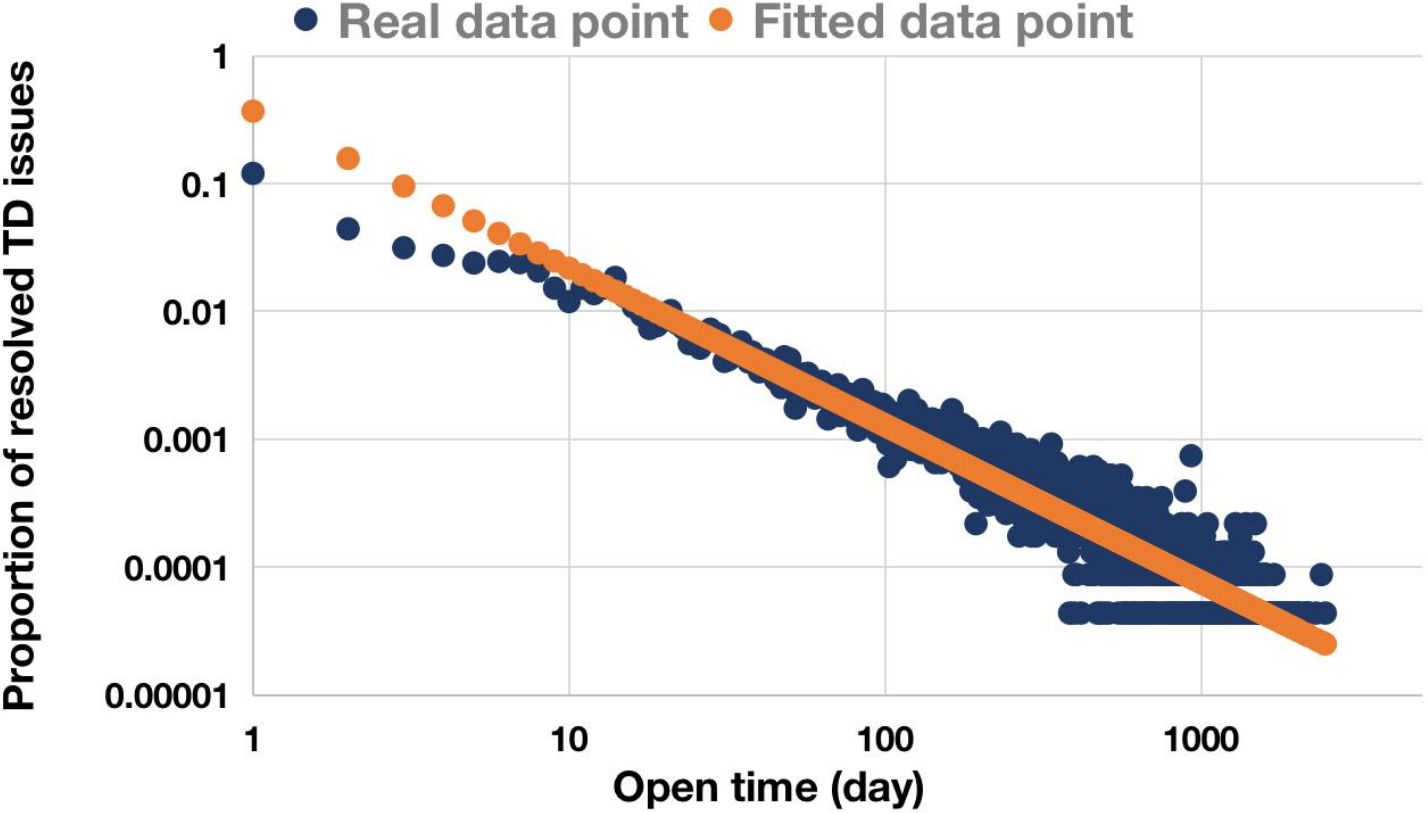}
    \caption{Distribution of the proportion of resolved TD issues over open time (RQ4).}
    \label{fig:dist_opentime}
\end{figure}

Out of all the 22,841 resolved TD issues, 2,761 (12.1\%), 1,012 (4.4\%), and 7,601 (33.3\%) were resolved in one, two, and less than ten days, respectively. The median is 25 days. As shown in Figure \ref{fig:dist_opentime}, the distribution of the proportion of resolved TD issues over open time (day) follows a power law \citep{BaAl1999}, \emph{i.e.}, $y=bx^{-a}$, where $y$ denotes the ratio of resolved TD issues, $x$ denotes the open time of the resolved TD issue, $a$ and $b$ are constants. The fitting results are $a=1.228$ and $b=0.369$, \emph{i.e.}, $y=0.369x^{-1.228}$ (see Figure \ref{fig:dist_opentime}). 

\begin{figure}[htpb]
    \centering
    \includegraphics[width=8cm]{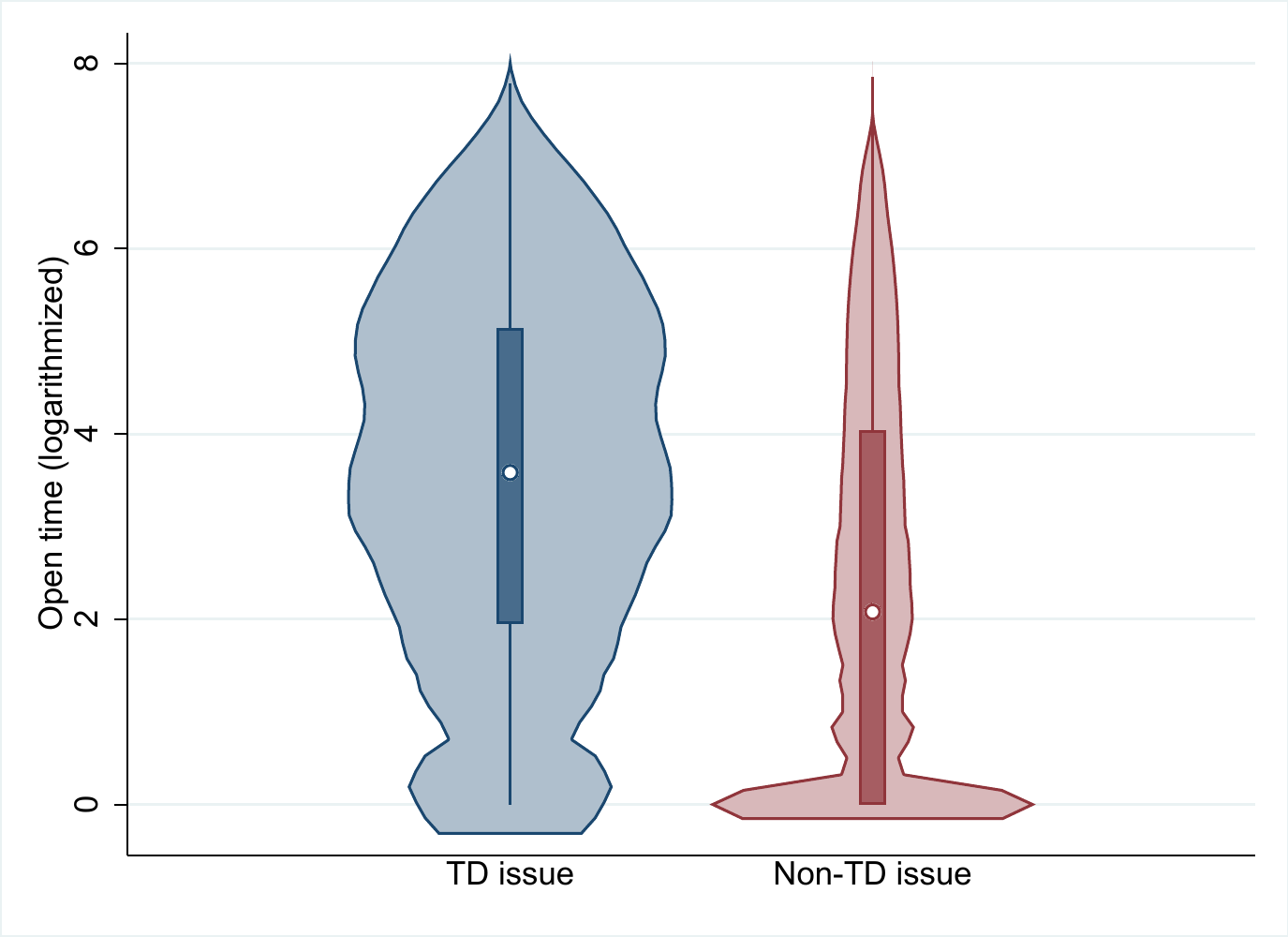}
    \caption{Distribution of the open time of resolved TD issues (RQ4).}
    \label{fig:opentime_violin}
\end{figure}

In addition, in order to study the difference on the open time between the resolved TD issues and non-TD issues, we calculated their distributions, which show that: 
(1) There are discrete values of the distribution of open time for the TD issues and non-TD issues, and the distributions show a right skewed pattern. The maximum open time for the TD issues is 2,403 days, which is extremely long.
(2) The inter-quartile range shows that the open time of TD issues is longer than that of non-TD issues. The average open time of resolved TD issues is 84.5\% longer than that of resolved non-TD issues, which is 155 days for the former and 84 days for the latter.
(3) A vast majority of TD issues are resolved in a long period of time compared to non-TD issues.
Figure \ref{fig:opentime_violin} provides a violin plot based on the distributions of resolved TD issues and non-TD issues. The vertical coordinate of this plot is logarithmized (into natural logarithm) in order to ensure that the readers' attention can be focused on the core data (\textit{i.e.}, the inter-quartile range).

\subsubsection{Characteristics of resolved TD issues}

The results of Mann-Whitney U tests on the characteristics of resolved TD issues and non-TD issues are shown in Table \ref{table:p-value}, in which $Avg_{TD}$ and $Avg_{non-TD}$ denote the average values of the characteristic in the corresponding row of resolved TD issues and non-TD issues, respectively.

It can be found that for IssueSize and ParticipantNo, their \textit{p-values} are less than 0.05. Considering the average values of these two characteristics, it means that issue characteristics IssueSize and ParticipantNo of resolved TD issues are significantly smaller than those of resolved non-TD issues, respectively.  
On the contrary, the \textit{p-values} of CmntSize and CmntNo are larger than 0.05, which means that there is no significant difference between resolved TD issues and non-TD issues with respect to characteristics CmntSize and CmntNo.

\subsubsection{Reopen of TD issues}
There are 24 repositories with at least 100 resolved TD issues. Out of the total 5,202 resolved TD issues in the 24 repositories, 326 (6.3\%) resolved TD issues had been reopened before. In addition, for 205,927 resolved non-TD issues, 9,677 (4.7\%) resolved non-TD issues had been reopened before. 
From the percentages mentioned above, \textbf{it can be concluded that TD issues are more likely to be fixed for multiple times}.

\begin{table}[]
\centering
\caption{Results of comparison on issue characteristics (RQ4).}
\scalebox{1.00}{
\begin{tabular}{lrrr}
\hline
\textbf{Feature} & \textbf{$Avg_{AR}$} & \textbf{$Avg_{AR}$} & \textit{\textbf{p-value}} \\ \hline
IssueSize      & 723.22      & 914.66        & 0.000      \\
CmntNo         & 3.87        & 3.87          & 0.056 \\
CmntSize       & 991.99      & 1,037.50       & 0.492 \\ 
ParticipantNo  & 2.13        & 2.27          & 0.006 \\ \hline
\end{tabular}
}\label{table:p-value}
\end{table}

\begin{center}
\fbox{\parbox{0.475\textwidth}{\textbf{Summary (RQ4)}: All TD issues were resolved in 31.6\% of the repositories while no TD issues were resolved at all in 27.3\% of the repositories. The distribution of the proportion of resolved TD issues over open time obeys a power law. The mean open time of resolved TD issues is 84.5\% longer than that of resolved non-TD issues. TD issues are more likely to be reopened than non-TD issues. }}
\end{center}


\subsection{RQ5: Continuity of TDM}
\subsubsection{Repositories with different TDM continuity levels} After calculation, the value of $Period_{Y}$ is around 368 and the value of $Count_{Z}$ is 8. Following the definitions of the three continuity levels, 1,187 (32.9\%) repositories with TD labels are identified as Abandoned, while only 298 (8.2\%) repositories with TD labels are identified as Consistent.
The number of repositories of level Abandoned is around 4 times as large as the number of those of level Consistent. 
In addition, the remaining repositories are identified as Unclear. Since too few TD issues are managed or the period for TDM is too short, it is not clear whether these repositories are consistently using or have abandoned TDM.

\subsubsection{Characteristics of repositories with different TDM continuity levels} We compared the characteristics of the repositories of level Consistent and repositories of level Abandoned, and the specific results of Mann-Whitney U tests on the 7 characteristics of the repositories of level Consistent or Abandoned are shown in Table \ref{table:rq5-p-value}, in which $Avg_{CR}$ and $Avg_{AR}$ denote the average values of the characteristic in the corresponding row of the repositories of level Consistent and repositories of level Abandoned, respectively. We found that the repositories of these two levels differ significantly within the 95\% confidence interval on 6 characteristics: RepoIssueNo, RepoResolvedIssuePro, RepoTDRepoterNo, RepoTDIssueNo, RepoOpentimeAvg, and RepoCmntSizeAvg. Specifically, the average of these 6 characteristics are higher for repositories of level Consistent than that for repositories of level Abandoned.

\begin{table}[]
\caption{Results of comparison
on repository characteristics (RQ5).}
\scalebox{1.00}{
\begin{tabular}{lrrr}
\hline
\textbf{Characteristic} & \textbf{$Avg_{AR}$} & \textbf{$Avg_{CR}$} & \textbf{\textit{p-value}} \\ \hline
RepoIssueNo      & 242.206             & 1765.034             & \textless{0.001}            \\
RepoResolvedIssuePro     & 0.734               & 0.836               & \textless{0.001}            \\
RepoTDRepoterNo  & 1.832               & 8.028               & \textless{0.001}            \\
RepoTDIssueNo   & 7.234              & 53.652              & \textless{0.001}            \\
RepoResolvedTDIssuePro   & 0.603               & 0.634               & 0.872            \\
RepoOpentimeAvg  & 113.941              & 165.543             & \textless{0.001}            \\
RepoCmntSizeAvg  & 346.598             & 521.687             & \textless{0.001}           \\ \hline
\end{tabular}
}\label{table:rq5-p-value}
\end{table}

\begin{center}
\fbox{\parbox{0.475\textwidth}{\textbf{Summary (RQ5)}: Surprisingly, only 8.2\% (\textit{i.e.}, 298) of the repositories with TD labels adopt TDM consistently, while 32.9\% (\textit{i.e.}, 1,187) of the repositories with TD labels have abandoned TDM. The repositories with consistent TDM have significantly more issues, reporters of TD issues, TD issues, and open time as well as larger size of comments and a higher proportion of resolved issues than the repositories with abandoned TDM.}}
\end{center}


\section{Discussion}\label{chap:discussion}
\subsection{Interpretation of Study Results}
\textbf{RQ1:} 
First, the lasting and fast increase of the number of repositories with TD labels and TD issues indicates that the TD concept has been increasingly popular over the last decade. In other words, the OSS community on GitHub has been raising the awareness of TDM in software development. 
Second, the power-law characteristic of the distribution of ratio of repositories with TD labels over the number of TD issues indicates that only a small portion of repositories have a large number of TD issues and a large portion of repositories have only a few TD issues.

\textbf{RQ2:} The results of our manual analysis on sampled TD instances indicate that 1) both internal QAs (\textit{e.g.}, maintainability) and external QAs (\textit{e.g.}, usability) are affected by TD, 2) internal QAs (maintainability in particular) are affected the most by TD. In addition, co-occurring labels with TD labels are extremely diverse and can carry different kinds of messages. They may help to identify the affected QAs, TD types, purposes, and so forth.

\textbf{RQ3:} First, around a half of all the TD issues were identified when they were reported, which indicates that the reporters of such issues may have good awareness of TDM and that such issues might be relatively easy to be identified as TD. 
Second, in 91.9\% of the repositories, all TD issues were identified only by no more than four participants, which means that only a really small number of participants put TDM into practice in most repositories adopting TDM.
Third, the fact that 80.5\% of the TD issues were identified by issue reporters indicates that issue reporters tend to pay more attention to the characteristics of issues.
More than a half of the repositories with TD labels adopt TDM in the early phase of the development lifecycle, which indicates an early awareness of TDM in most repositories with TD labels.

\textbf{RQ4:} First, TD issues did not receive enough attention considering that the median of the proportion of resolved TD issues for all repositories is 0.582 and no TD issues were resolved in more than a fourth of all repositories. Second, the power-law characteristic of the distribution of the proportion of resolved TD issues over open time indicates that most resolved TD issues experienced short open time and a small portion of resolved TD issues had long open time. Third, TD issues take much (\textit{i.e.}, 84.5\%) longer to be resolved than non-TD issues. One potential reason is that most TD issues are not bugs and invisible to users, and thus they are usually assigned with a relatively low priority. Fourth, on average, a non-TD issue 
attracts more participants to join discussion than a TD issue. In other words, a non-TD issue receives more attention than a TD issue. Finally, on average, a TD issue is more likely to be reopened than a non-TD issue, which means that TD issues are more difficult to be cleanly fixed than non-TD issues.

\textbf{RQ5:} Over the 3,598 repositories with TD labels, around a third (32.9\%) of the repositories have abandoned TDM, which is partially due to the situation that only one participant is involved in TDM for 69.0\% of all the repositories. If this only participant left the repository, the practice of TDM there is likely to be dropped. This is further confirmed by that the number of TD issue reporters for the repositories keeping consistent TDM is significantly larger than that for the repositories having abandoned TDM.

\subsection{Implications}
Based on the study results and their analysis, we obtained a number of implications for practitioners and researchers, respectively.
\subsubsection{Implications for practitioners} 

(1) The OSS community are embracing the TD concept, which in turn reflects the practical value of TDM. The TD concept is well adopted in some large repositories, such as \textit{microsoft/vscode}, in the sense that hundreds and even more of TD issues were lastingly managed in such repositories.

(2) Although the TD concept has been adopted in some OSS repositories, TDM is possible to be ignored in the sense that no TD issues were resolved in 27.3\% of the repositories, and TDM can even be abandoned in the sense that 32.9\% of the repositories are identified as having abandoned TDM.  
In the context of OSS development, to benefit from TDM, we recommend to pre-define TD labels for a project (or repository) and to explicitly make clear and concrete TDM policies that the participants should obey. In OSS development, if there are no TDM policies exerted on a project, the awareness of TDM is likely to be brought away from the project by the participants who are the only ones care about TDM when they quit the project. 

\subsubsection{Implications for researchers} 

(1) As the TD practice has been increasingly popular in the OSS community, which resonates with the intensive research on TD in the last decade.   

(2) The reasons for why only one TD issue was identified in  31.1\% of all the repositories and only one participant identified all TD issues in 69.0\% of all the repositories, need to be further investigated. This can help us to understand why the TD concept is not adopted by other participants in such repositories, to evaluate the value of the TD concept in such repositories, and to explore the conditions required by well adoption of the TDM practices in such repositories.

(3) There is in need for a deep investigation on the cost/benefit of adopting TDM in OSS development, in order to clarify the reasons why around a third (32.9\%) of the repositories have abandoned TDM and only 8.2\% of the repositories keep TDM as a consistent practice.

(4) We highlight the gap between the academic research and OSS development practice on TDM. TDM has been a hot research topic in software engineering for more than a decade \citep{LiAvLi2015,CiLeMa2021}, a large number of research papers on TDM have been published. Although the awareness of TDM has been increasing in the sense that the growth of the number of repositories newly adopting TDM is higher than that of newly created repositories, there are only 298 repositories keep TDM as a consistent practice on the whole GitHub. This demonstrates a big gap between academic research on TDM and practical use of TDM in OSS development.


\section{Threats to Validity}\label{chap:threats}
There are several threats to the validity of the study results. We discuss these threats according to the guidelines in \citep{RuHo2009}. Internal validity is not discussed, since we did not study causal relationships.

\subsection{Construct validity}
Construct validity is concerned with whether the values of the variables (listed in Table \ref{table:dataitem}) we
obtained are in line with the real values that we expected.
First, considering that many projects hosted on GitHub do not record their issues on GitHub, our collected dataset is not complete. 
Second, it is possible that the keywords used to search TD issues do not cover all labels for tagging TD issues. However, to reduce this risk, we chose the string "technical debt" as well as its variable forms (\textit{e.g.}, word abbreviations), and selected the appropriate TD types according to the study of Li \emph{et al}. \citep{LiAvLi2015}. Admittedly, issues with the list of TD labels may not cover all issues containing TD. However, in order to investigate the practice of TDM performed by practitioners in the OSS development, we do not want to judge by ourselves whether an issue contains TD or not, but let the practitioner speak out. Therefore, we chose the labels that are directly related to TD as TD labels.
Third, the 914 TD issues without labeled events may influence the completeness of the data. Dataset DS1 excluding 914 TD issues without labeled events was analyzed to partially answer RQ3. Considering that the 914 TD issues did not come from a small number of repositories and were not reported in a narrow time slot, the threat caused should be limited. 
Finally, the results of our manual analysis of the affected QAs and TD types of TD issues may be biased due to the differences of experience and knowledge between the researchers. This threat was greatly mitigated by the pilot tagging process, discussion of the disagreement of individual tagging results with a third researcher.

\subsection{External validity}
External validity is concerned with the generalizability of
the study results. We only investigated issues reported on repositories hosted on GitHub, therefore, the findings obtained in this paper are merely valid for GitHub, and may not be generalized to other OSS platforms or ecosystems. However, GitHub is the largest OSS platform, the repositories included in this study come from different application domains and adopt a wide range programming languages, and thus our datasets used in this study should be sufficiently representative. Hence, the threats to external validity should be alleviated.  

\subsection{Reliability}
Reliability is concerned with whether the study yields the same
results when it is replicated by other researchers.
In all the data analysis tasks, only descriptive statistics were used, and all datasets are provided on GitHub. Hence, the threats to reliability should be minimal.

\section{Conclusions and Future Work}\label{chap:conclusions}
In this work, we conducted an empirical study on all issues with TD labels on the whole GitHub to investigate the execution of TDM in OSS development. The following findings were obtained: 
(1) the awareness of TDM in the OSS community has been arising; 
(2) maintainability, reliability, and functional suitability are the top three QAs affected most by TD issues, and design TD, architectural TD, and test TD are the top three TD types occurred most frequently in TD issues; 
(3) only one TD issue was identified in 31\% of the repositories, all TD issues were identified by only one developer in 69\% of the repositories, 81\% of TD issues were identified by the issue reporters, 49\% of TD issues were identified when the issues were reported and the other 51\% of TD issues were identified during the resolution process of the issues; 
(4) TDM was ignored in 27.3\% of the repositories after TD issues were identified; and (5) only 8.2\% of the repositories keep TDM as a consistent practice, while 32.9\% of the repositories have abandoned TDM.

Based on the findings in this work, we plan (1) to conduct a broad survey on TDM in the closed source software industry to understand the state of the adoption and execution of TDM, and (2) to investigate the reasons for ignoring and abandoning TDM in the repositories that have adopted TDM. 
\section*{Acknowledgment}
This work is supported by the Natural Science Foundation of Hubei Province of China under Grant No. 2021CFB577, and the National Natural Science Foundation of China under Grant No. 62176099 and No. 62172311.

\printcredits

\bibliographystyle{cas-model2-names}
\bibliography{references}

\balance
\end{sloppypar}
\end{document}